\documentclass[journal]{IEEEtranTIE}
\usepackage{graphicx}
\usepackage{subfigure}
\usepackage{cite}
\usepackage{picinpar}
\usepackage{amsmath}
\usepackage{amssymb}
\usepackage{mathrsfs}
\usepackage{url}
\usepackage{flushend}
\usepackage[latin1]{inputenc}
\usepackage{colortbl}
\usepackage{soul}
\usepackage{multirow}
\usepackage{pifont}
\usepackage{color}
\usepackage{alltt}
\usepackage[hidelinks]{hyperref}
\usepackage{enumerate}
\usepackage{siunitx}
\usepackage{breakurl}
\usepackage{epstopdf}
\usepackage{pbox}
\usepackage{CJK}
\usepackage{bm}
\usepackage{makecell}

\begin{document}
\title{Deep Learning based Intelligent Coin-tap Test for Defect Recognition}

\author{
	\vskip 1em
	
	Hongyu Li, 
	Peng Jiang, 
	and Tiejun Wang*

	\thanks{
		This work was supported in part by the National Science and Technology Major Project (2019-VII-0007-0147), in part by the National Natural Science Foundation of China (11902240), and in part by the fund of Innovative Scientific program of CNNC.
		
		H.Y. Li, P. Jiang, and T.J. Wang* (the corresponding author) are with the State Key Lab for Strength and Vibration of Mechanical Structures, Department of Engineering Mechanics, Xi'an Jiaotong University, Xi'an, 710049, China (e-mail: lihongyu@stu.xjtu.edu.cn, jiangpeng219@mail.xjtu.edu.cn, 973wtj@xjtu.edu.cn).
	}
}

\maketitle

\begin{abstract}
The coin-tap test is a convenient and primary method for non-destructive testing, while its manual on-site operation is tough and costly.
With the help of the latest intelligent signal processing method, convolutional neural networks (CNN), we achieve an intelligent coin-tap test which exhibited superior performance in recognizing the defects.
However, this success of CNNs relies on plenty of well-labeled data from the identical scenario, which could be difficult to get for many real industrial practices.
This paper further develops transfer learning strategies for this issue, that is, to transfer the model trained on data of one scenario to another.
In experiments, the result presents a notable improvement by using domain adaptation and pseudo label learning strategies.
Hence, it becomes possible to apply the model into scenarios with none or little (less than 10\%) labeled data adopting the transfer learning strategies proposed herein.
In addition, we used a benchmark dataset constructed ourselves throughout this study.
This benchmark dataset for the coin-tap test containing around 100,000 sound signals is published at \url{https://github.com/PPhub-hy/torch-tapnet}.
\end{abstract}

\begin{IEEEkeywords}
Intelligent defect recognition, non-destructive testing, coin-tap test, convolutional neural networks, deep transfer learning.
\end{IEEEkeywords}

\definecolor{limegreen}{rgb}{0.2, 0.8, 0.2}
\definecolor{forestgreen}{rgb}{0.13, 0.55, 0.13}
\definecolor{greenhtml}{rgb}{0.0, 0.5, 0.0}

\section{Introduction}

Non-destructive testing (NDT) plays a key role in assessing the quality of industrial structures and components, for example, nuclear power structures, weld lines, pipes and pressure vessels, fiber-reinforced composite structures, etc.
The coin-tap test, also called impact echo (IE)~\cite{Jaeger1996DETECTINGVI} or local acoustic resonance spectroscopy (LARS)~\cite{Jatzlau2020LocalAR}, is one of the most widely-used, convenient and primary methods of NDT.
During the coin-tap test, free vibration is excited around the tap points by tapping, and then, the vibrational frequency in the areas containing defects will be quite different from that in the intact areas.
This is because the local stiffness changes owing to the presence of damage such as cavities or hard/soft inclusions~\cite{Wang1991ACD,Wang1992UnifiedCM}.
Vibration produces sounds, thus, the acoustic signals can indicate anomalies caused by various defects inside structures.
This method is used in detecting hollows, delamination in infrastructures, and foreign substances, bubbles, debonding inside the laminates~\cite{Jaeger1996DETECTINGVI,Jatzlau2020LocalAR}.
While, in practice, it is inefficient to conduct large-scale coin-tap tests by human ears, and it is costly to train journeymen~\cite{Andreisek2016TheVT}.
Thus, an intelligent coin-tap test is urgent and significant, while it is also a meaningful attempt towards intelligent non-destructive testing (iNDT).

The key for an intelligent coin-tap test is the accurate pattern recognition of vibration/acoustic signals.
For this problem, conventionally, the patterns were recognized according to manually extracted features (such as skewness, kurtosis, etc.) of signals by spectrum analysis like the wavelet package decomposition (WPD)~\cite{Lei2007FaultDO}.
For the coin-tap test, full duration at half maximum (FDHM) of the force signal is the most commonly used feature~\cite{Jatzlau2020LocalAR}.
On this basis, by using machine learning technology, some data-driven methods can recognize the signals without expert knowledge~\cite{Jia2016DeepNN, Pan2018LiftingNetAN, He2021DeepVA}.
However, the manually extracted features are not sufficient to achieve robust defect detection for real applications.
Therefore, convolutional neural networks (CNN)~\cite{Zhang2018ADC, Jiang2019MultiscaleCN, Zhao2019MultipleWC, Miao2021AnEM} including millions of trainable weights are used recently as intelligent approaches to extract features from the raw signal or its time-frequency-domain components for classification.
Notably, Dorafshan \textit{et al.}~\cite{Dorafshan2020DeepLM,Dorafshan2020EvaluationOB} conducted a successful application of deep learning technology on the IE test, i.e., the coin-tap test.
Their attempts are illuminating but still have plenty to work for, such as constructing better CNN architecture and benchmark dataset.

Take one more step forward to consider a further challenge that, it could be difficult to get plenty of well-labeled data from the identical scenario in many real industrial practices.
Applying models trained in similar scenarios could be a good way for solving this issue.
Therefore, we conduct studies taking the model transfer across different materials as the similar scenarios for example.
A special training technique, transfer learning, provides approaches for this problem, and it recently takes the cutting edge of intelligent signal analysis~\cite{Guo2019DeepCT, Song2020RetrainingSD, Yang2019AnIF, Li2019CrossDomainFD, Wang2019AHD, Han2019ANA, Yang2020APK, Li2020DeepLM, Jiao2020UnsupervisedAA, Ragab2021AdversarialMD}.
Most of the successes are for fault diagnosis of machinery, for example, transferring the models for diagnosis between induction motors and gearboxes~\cite{Song2020RetrainingSD}, transferring the models from laboratory bearing to locomotive bearings~\cite{Yang2019AnIF}, etc.
Among them, most papers adopt the idea of domain adaptation, and there are two different implementations.
One is statistical distribution adaptation.
Li \textit{et al.}~\cite{Li2019CrossDomainFD} and Wang \textit{et al.}~\cite{Wang2019AHD} used the correlation alignment (CORAL) loss for cross-domain network training.
Impressively, Yang \textit{et al.}~\cite{Yang2020APK} tried polynomial kernels when calculating the inner product of samples in reproduced kernel Hilbert space (RKHS) for calculating the maximum mean discrepancy (MMD).
The other is adversarial network training.
Jiao \textit{et al.}~\cite{Jiao2020UnsupervisedAA} utilized this method based on the multi-classifier discrepancy.
Han \textit{et al.}~\cite{Han2019ANA} and
Ragab \textit{et al.}~\cite{Ragab2021AdversarialMD} improved the models via confrontation between the feature extractor and an auxiliary discriminator.
Hence, these advanced technologies could be effective for dealing with the data shortage in the practices of the intelligent coin-tap test herein.

This paper aims to develop an intelligent coin-tap test for defect recognition based on deep learning.
On the one hand, CNNs are trained to recognize the defect in scenarios with sufficient well-labeled data, on the other hand, we carry out a study about model transfer to extend the application to situations without sufficient labeled data.
Specifically, the models are transferred across different materials, and domain adaptation and pseudo label learning strategies are adopted to improve the accuracy.
This paper is organized as follows.
Section 2 describes the pattern recognition tasks concerned in this study.
Section 3 introduces the methodology we adopted and proposed.
In section 4, the process of data collection and the benchmark dataset constructed for the coin-tap tests are introduced.
Section 5 presents experiments, results, and discussions about the proposed deep learning methods.
In Section 6, conclusions and remarks are presented.
To help future works and reproduction of this study, we publish the full dataset, codes, and a quick demo at \url{https://github.com/PPhub-hy/torch-tapnet}.

\section{Statement of the problem}

Towards an intelligent coin-tap test, there are two tasks: i) the pattern recognition of signals for different defects, ii) transfer the model to unseen scenarios without sufficient labeled data.
The following describes them technically.

\subsubsection{Pattern recognition of defect}
Defect recognition via the coin-tap test is to classify the signals into patterns that correspond to different defects.
Feature extraction is a major part of this task.
As shown in Fig.~\ref{problemA}, we construct machine learning method $\mathfrak{L}$, that is, to map samples from the input space $\mathcal{X}$ to the feature space $\mathcal{F}$.
Then, there could be a clear boundary between samples of different classes in $\mathcal{F}$, therefore, it is easy to get them classified.

\subsubsection{Transfer learning via domain adaptation}
In practice, we sometimes need to apply models into unseen scenarios without sufficient label samples (for example, another material).
Transfer learning is the strategy to transfer the knowledge learned from the training data to the target scenario.
Domain adaptation is the special trick we adopted.
The word 'domain' means the application scenario, i.e., the material in our study.
The domain is mathematically a distribution $\mathcal{D}$ in the input space $\mathcal{X}$.
The source domain $\mathcal{D}_s$ and the target domain $\mathcal{D}_t$ denotes the training scenario and the unseen scenario respectively.
A domain is characterized by a probability distribution, but the real scenarios are always complex and not analytical.
This means the probability density of the distribution can only be approximately represented by the density of samples in the space.
Specifically, $\mathcal{D}_s = \{(x^s_i,y^s_i)\}^{n_s}_{i=1}$ and $\mathcal{D}_t = \{x^t_i\}^{n_t}_{i=1}$ where the $(x^s_i,y^s_i)$ and $x^t_i$ are samples in the source and target domain respectively.
For the sake of convenience, we use $p^\mathcal{X}$ and $q^\mathcal{X}$ to represent the probability distributions of source domain and target domain in space $\mathcal{X}$.

\begin{figure}[t]
\centering
\subfigure[feature extraction and patter recognition]
    {
    \label{problemA}
    \includegraphics[width=0.85\linewidth]{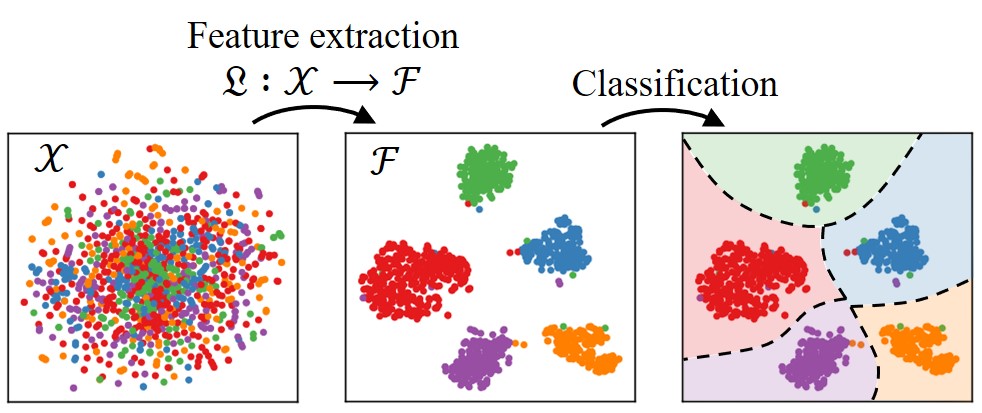}
    }
\subfigure[transfer learning via domain adaptation]
    {
    \label{problemB}
    \includegraphics[width=0.9\linewidth]{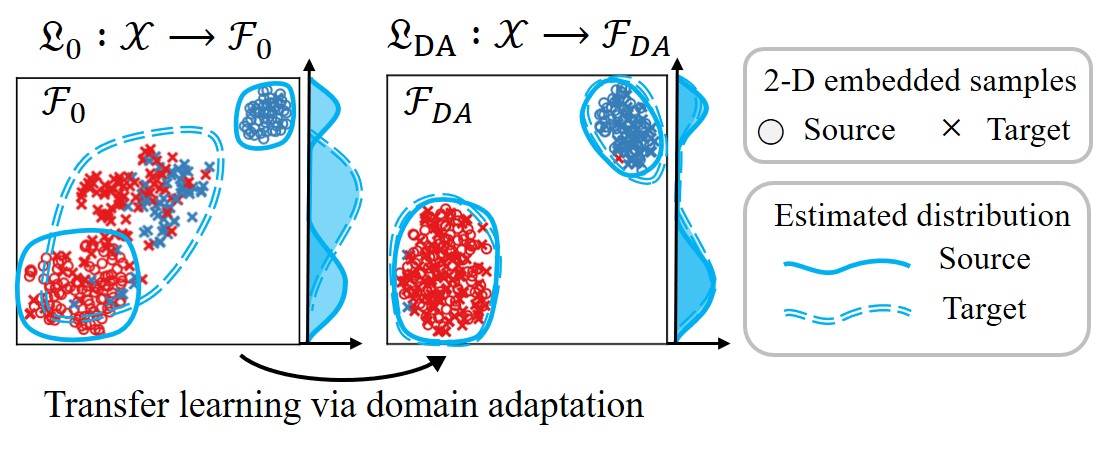}
    }

\caption{Illustration of the two signal processing tasks towards intelligent coin-tap test.
$\mathfrak{L}$, $\mathfrak{L}_0$, and $\mathfrak{L}_{DA}$ denotes deep learning algorithms for signal recognition, and $\mathcal{F}$, $\mathcal{F}_0$, and $\mathcal{F}_{DA}$ are the feature space generated by these algorithms respectively.
$\mathcal{X}$ is the input space. The subscript 'DA' denotes the algorithm with transfer learning technology.}
\label{fig:Illustration}
\end{figure}

The domain adaptation realizes transfer learning by adapting the distribution $p$ and $q$ when training the networks.
As shown in Fig.~\ref{problemB}, comparing to $\mathfrak{L}_0$ without domain adaptation, the goal of $\mathfrak{L}_{DA}$ is not only to recognize the patterns but also to narrow the gap between $p^\mathcal{F}$ and $q^\mathcal{F}$ when mapping the samples into the feature space $\mathcal{F}$.

Some principal assumptions are worth noting for practice:
i) The input/output spaces of the source and target domains are isomorphic, i.e., the dimensions and the meanings of the input/output in the two domains should be the same.
ii) The knowledge provided by the source domain has an intersection big enough with the target domain that to ensure the knowledge transfer valuable.

\section{Methodology}

\subsection{Mechanics principles of the coin-tap test}

The vibration/sound signals obtained during the coin-tap test are closely related to the local dynamic characteristics of the structures.
According to the classic theory by Cawley \textit{et al.}~\cite{Cawley1988THEMO}, the local stiffness is presented using an ideal model with two springs $k_c$ and $k_d$ respectively representing the local stiffness and the contact stiffness.
For a given tapping head and structure, $k_c$ is constant, whereas $k_d$ changes at the points with defects, which leads to changes in the modes of vibration and sound.
With the help of acoustic simulation and experiments, Groschup \textit{et al.}~\cite{Groschup2015MEMSMA} demonstrated the signals released after tapping more clearly.
In fact, since the shape and characteristics of real defects are unpredictable, it is very hard to identify the defects by these theories.
However, these theories guarantee that there is a strong relevance between the tapping sound and defects inside the structures, which is a key prerequisite for developing data-driven approaches to this problem.

\subsection{Proposed method based on deep learning}

Herein, we propose a deep learning based algorithm for recognizing the sound signals for the intelligent coin-tap test.
Specifically, it includes a CNN model $F(x,\Theta)$ and a series of training techniques for optimizing the weights $\Theta$.

\subsubsection{Deep convolutional neural networks}

\begin{figure}[t]
\centering
\subfigure[~\cite{Zhang2018ADC}]
    {
    \label{TICNN}
    \includegraphics[width=0.262\linewidth]{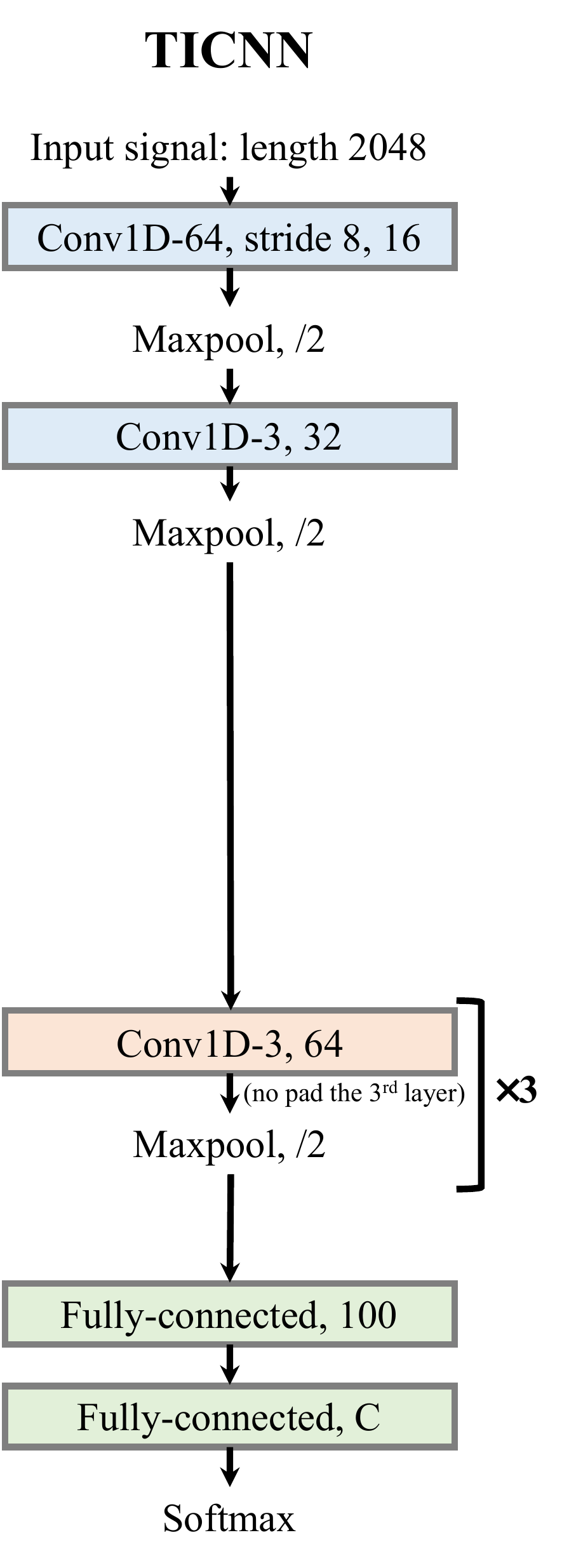}
    }
\subfigure[plain CNN]
    {
    \label{plainCNN}
    \includegraphics[width=0.27\linewidth]{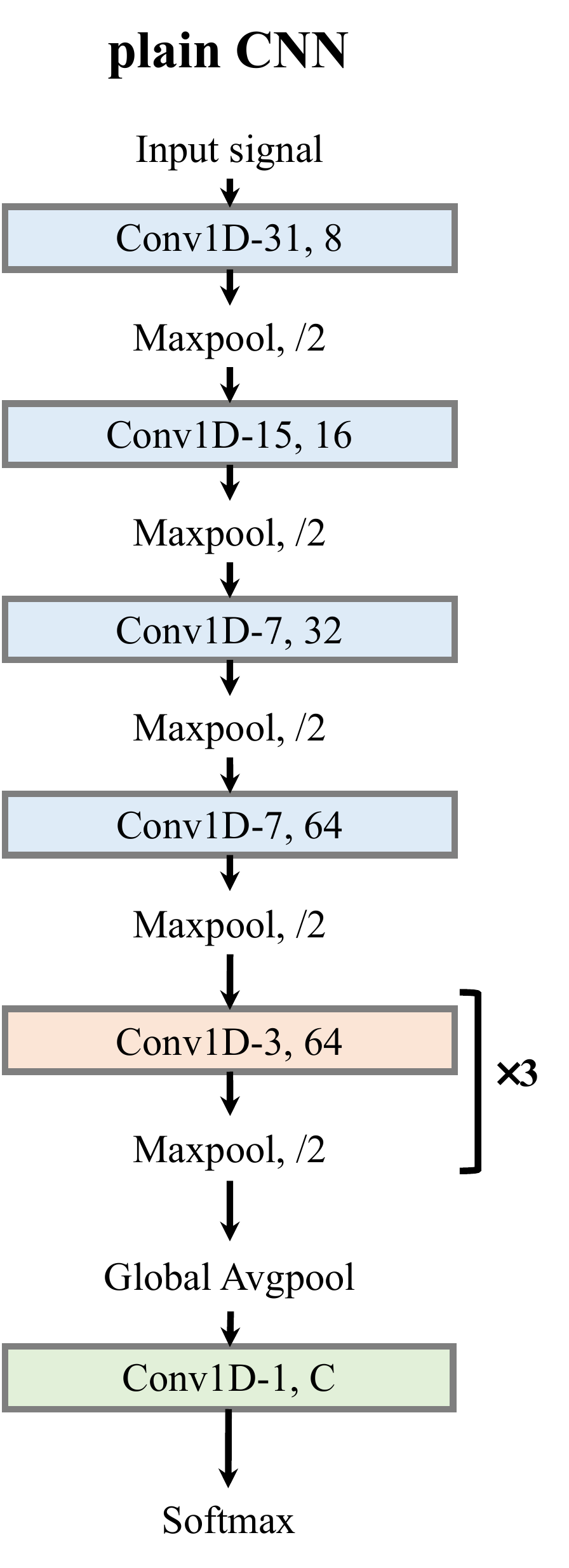}
    }
\subfigure[ResNet]
    {
    \label{resnet}
    \includegraphics[width=0.345\linewidth]{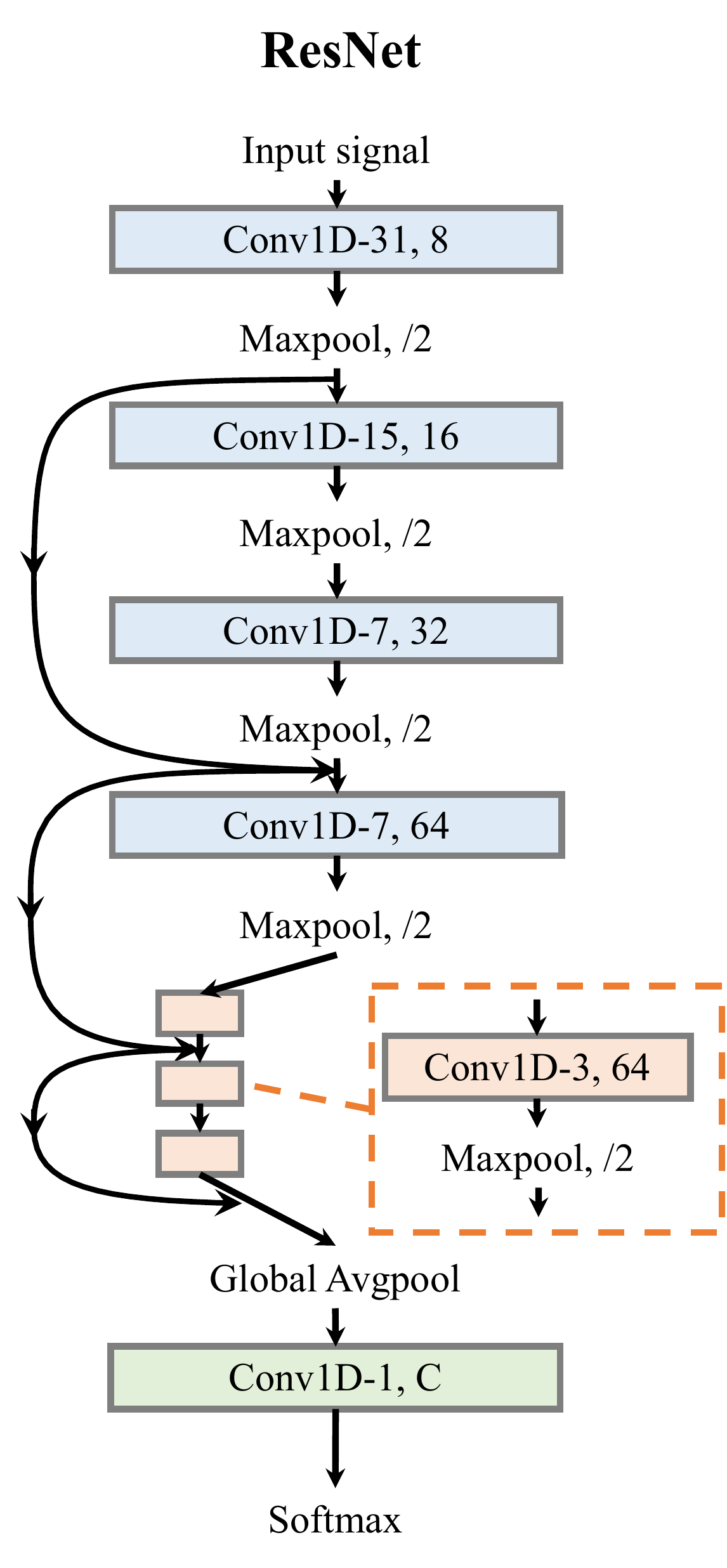}
    }
\subfigure[DenseNet]
    {
    \label{densenet}
    \includegraphics[width=0.95\linewidth]{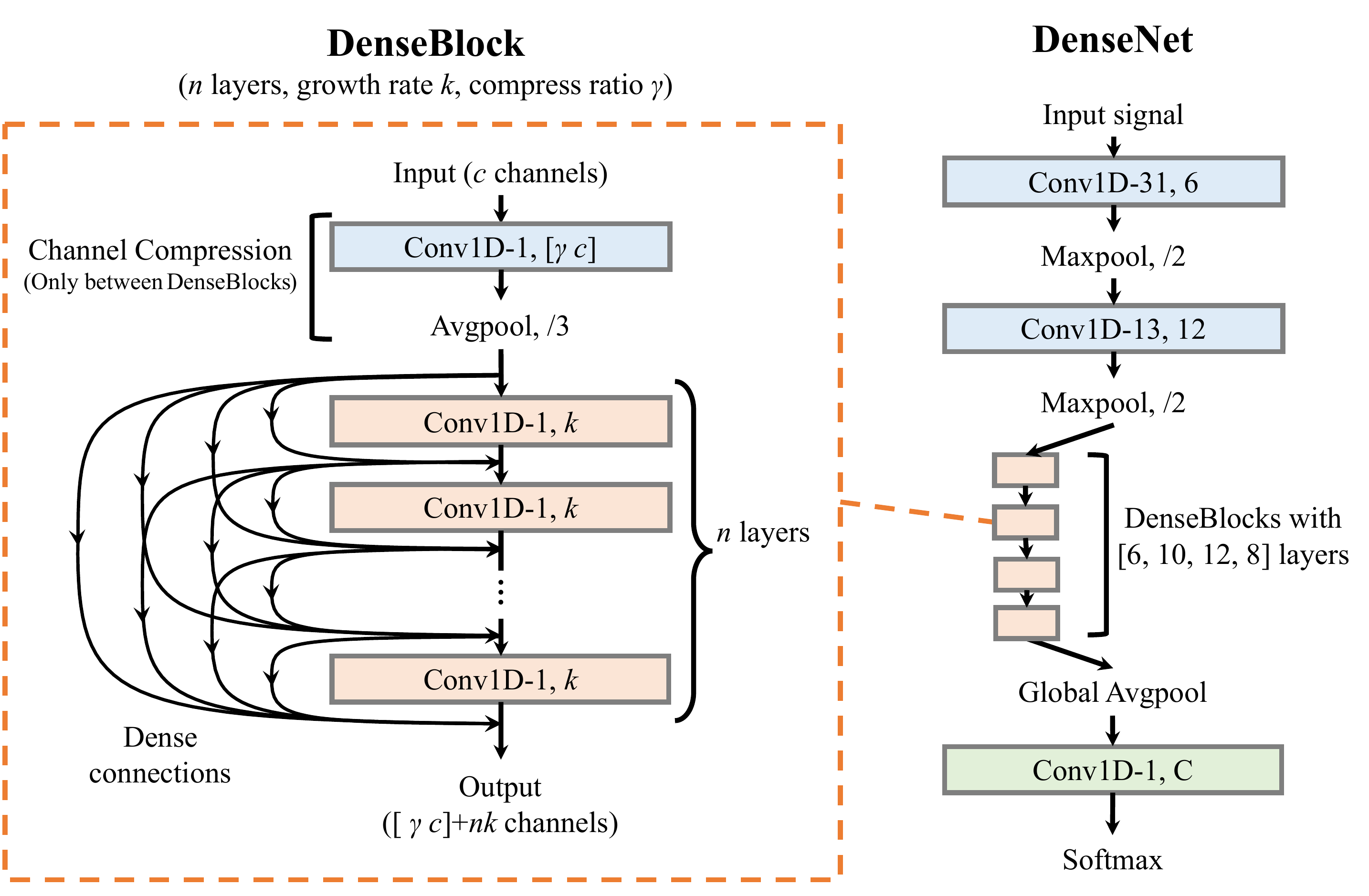}
    }
\caption{Overview of the CNNs constructed towards an intelligent coin-tap test. The convolutional layers are written as 'Conv1D-$s$, $c$' where $s$ is the kernel size and $c$ is the output channels. The $C$ in the last layer of each network denotes the number of classes.}
\label{fig:CNNs}
\end{figure}

The proposed deep CNN model $F(x,\Theta)$ is a long composite function of the layers $\{f^l\}$ to map the input $x$ to the feature space, i.e., $F=f^n\circ...\circ f^2 \circ f^1$ where n is the number of layers, and $\Theta$ represents the set of learnable parameters of $F$.
Following typical ideas, we construct three CNNs as the function $F$ to achieve intelligent signal processing for the coin-tap test.
Fig.~\ref{fig:CNNs} shows a sketch of these CNNs comparing with the architecture constructed by Zhang \textit{et al.}~\cite{Zhang2018ADC}.

Plain CNN is the most classic one, which comprises a series of stacked convolutional layers and pooling layers.
A sequence of batch normalization and nonlinear function for activating (the BN-ReLU) is inserted between every two convolutional layers.
One BN-ReLU-Conv block is mathematically expressed as:
\begin{align}
\bm{Z}^{l+1} = f^l(\bm{Z}^l) = Conv \circ ReLU \circ BN(\bm{Z}^l),
\end{align}
where $\bm{Z}^{l}$ denotes the input tensor of $l^{th}$ layer.

Network for residual learning is now popular and firstly proposed by He \textit{et al.}~\cite{He2016DeepRL} to train extremely deep CNNs (up to 1k+ layers).
Following this idea, our second network (ResNet) is constructed by adding identical mappings between layers based on the plain CNN.
Mathematically, it is expressed as:
\begin{align}
\begin{split}
\left \{
\begin{array}{ll}
\bm{Z}^{l+1} = f^l(\bm{Z}^l) \\
\bm{Z}^{l+2} = f^{l+1}(\bm{Z}^{l+1}) + pad(\bm{Z}^l)
\end{array}
\right.
(l=1,3,5,...),
\end{split}
\end{align}
where $pad(\bm{Z}^l)$ means to pad the tensor $\bm{Z}^{l+1}$ to match the size of $f(\bm{Z}^{l+2})$.

The densely connected network (DenseNet)~\cite{Huang2017DenselyCC} is reported to perform better than the ResNet.
Following this idea, we construct DenseNet with four dense blocks for recognizing the vibration signals.
For layers inside dense blocks, the operation is expressed as:
\begin{align}
\bm{Z}^{l+1} = f^l([\bm{Z}^i]^l_{i=0})
\end{align}
where the $[\bm{Z}^i]^l_{i=0}$ means to concatenate all the former tensors in the current block.
Regarding the hyper-parameters used in DenseNet, we set the output channel of each layer $k$ as 6 and the reduction ratio $\gamma$ as 0.5.

\subsubsection{Training for classification}
The training process is to optimize the weights $\{\theta_i| \theta_i \in \Theta\}$ in $F(x,\Theta)$ according to a predefined target, i.e., the loss function $\mathcal{L}$.
In specific, we get $\frac{\partial \mathcal{L}}{\partial \theta_i}$ via back propagation, and then update each $\theta_i$ by an optimizer as one iteration.

The primary part of the loss function is for classification.
After using the Softmax nonlinear activation, the output of one sample $x_i$ from the model $F$ is $y_i = \{p_{ic}\}_{c=1}^C$ where the components of $y_i$ denote a normalized probability that indicates the classes of the input sample.
We use cross entropy as the classification loss: $\mathcal{L}_c = -\frac{1}{N}\sum_{i=1}^N\sum_{c=0}^{C}\hat{p}_{ic}log(p_{ic})$, where $N$ is the number of input samples, and $p_{ic}$ is the probability component of class $c$ of the $i^{th}$ input sample.
$\hat{p}_{ic}$ is the real label.
It equals 1 if the $i^{th}$ sample exactly belongs to class $c$, it is 0 otherwise.
For direct classification without transfer learning, the loss function is $\mathcal{L} = \mathcal{L}_c + \lambda L_1$ where the $L_1$-loss is the weight decay used for regularization, and $\lambda$ is set to $10^{-4}$.

\subsubsection{Transfer learning via domain adaptation and pseudo label learning}

Domain adaptation aims to make the probability distribution $p^\mathcal{F}$ and $q^\mathcal{F}$ as similar as possible.
We use MMD to represent the discrepancy between two distributions.
The MMD inside the RKHS is represented as follows~\cite{Gretton2012AKT,Yang2020APK}:
\begin{align}
D_{\mathcal{H}}(p^\mathcal{F},q^\mathcal{F}) = \sup_{\phi(\cdot) \in \mathcal{H}}(E_{\hat{x}^s\sim p^\mathcal{F}}[\phi(\hat{x}^s)] - E_{\hat{x}^t\sim q^\mathcal{F}}[\phi(\hat{x}^t)]),
\label{fun:MMDt}
\end{align}
where $\hat{x}^s$ and $\hat{x}^t$ denote the samples that have been transformed to the feature space $\mathcal{F}$.
$\phi(\cdot)$ is the mapping operator to RKHS.
For practice, the MMD can be estimated as $\tilde{D}^2_{\mathcal{H}}$ by one mini-batch of samples $\{\hat{x}^s_i, \hat{x}^t_j\}_{i,j=1}^{n_s,n_t}$~\cite{Gretton2012AKT,Yang2020APK}:
\begin{align}
\begin{split}
\tilde{D}^2_{\mathcal{H}}(p^\mathcal{F},q^\mathcal{F}) &= \frac{1}{n_s^2}\sum_{i=1}^{n_s}\sum_{j=1}^{n_s}k(\hat{x}^s_i, \hat{x}^s_j) + \frac{1}{n_t^2}\sum_{i=1}^{n_t}\sum_{j=1}^{n_t}k(\hat{x}^t_i, \hat{x}^t_j)\\
&- \frac{2}{n_sn_t}\sum_{i=1}^{n_s}\sum_{j=1}^{n_t}k(\hat{x}^s_i, \hat{x}^t_j),
\label{fun:MMD}
\end{split}
\end{align}
where $k(x_i,x_j)$ is the kernel function that is used to implicitly represent the RKHS that $k(x_i,x_j)=\left< \phi(x_i),\phi(x_j)\right>$.
Specifically, we adopt the 5-fold Gaussian kernel as
\begin{align}
k(x_i,x_j) = \sum_{i=-2}^2 exp\left( -\frac{\Vert x_i-x_j \Vert ^2}{2(2^i \sigma)^2}\right)
\label{fun:kernel}
\end{align}
where $2^i\sigma$ denotes the bandwidths of the Gaussian kernel.
Like it says, the $\tilde{D}^2_{\mathcal{H}}$ is used as the loss of domain adaptation $\mathcal{L}_{da}$.

Although there is no label for samples of the target domain, the network itself may generate a considerable number of correct labels after a period of training.
Some researches show that it will bring improvement by feeding these pseudo labels back to the network~\cite{Yang2019AnIF, Song2020RetrainingSD}.
Thus, we adopt the pseudo label learning strategy and calculate the loss of it similarly to the classification loss, i.e., $\mathcal{L}_{pll} = -\frac{1}{N}\sum_{i=1}^N\sum_{c=0}^{C}\tilde{p}_{ic}log(p_{ic})$, where $\tilde{p}_{ic}$ equals to 1 when $c = \mathop{\arg\max}_{k} p_{ik}$, it is 0 otherwise.

\begin{figure}[t]
\centering
\includegraphics[width=0.8\linewidth]{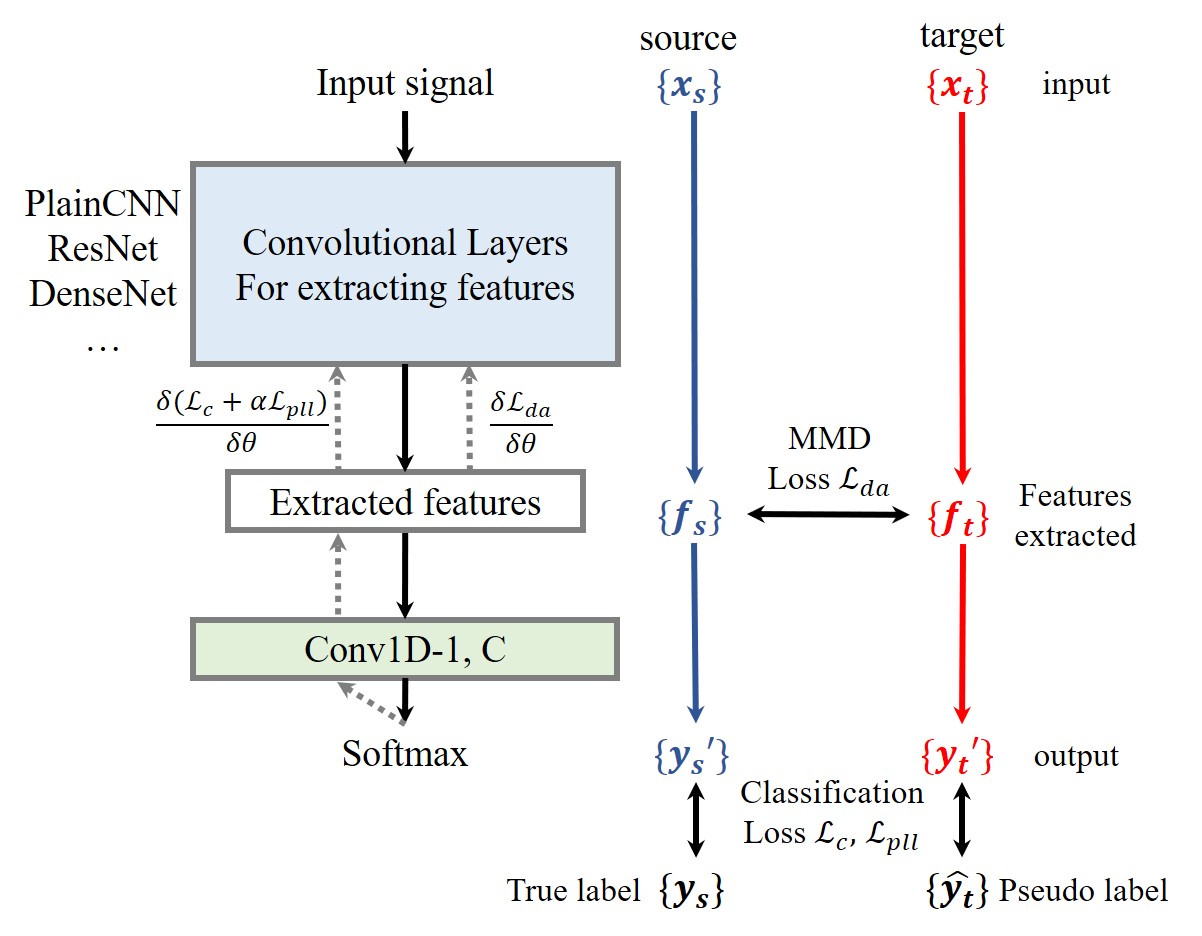}
\caption{Illustration of the training process of CNNs using domain adaptation. The left shows the structure of the CNN model, and the right shows the data flow during training. The gray dotted arrow is the back propagating flow of the gradient.}
\label{fig:Illu_DA}
\end{figure}

The overall loss function (the training objective) for domain transfer is written as a weighted sum of the three mentioned parts as
\begin{align}
\mathcal{L} = \mathcal{L}_c + \beta \mathcal{L}_{da} + \alpha \mathcal{L}_{pll} + \lambda L_1.
\end{align}
To be specific, we set $\beta=1$, and $\lambda$ (the weight decay) $=10^{-4}$.
Since the pseudo label learning should work after a period of training, $\alpha$ follows a segmented schedule as
\begin{align}
\begin{split}
\alpha=\left \{
\begin{array}{lc}
0 &t\leqslant0.1T \\
\frac{t-0.1T}{0.1T} &0.1T<t\leqslant0.2T \\
1 &t>0.2T \\
\end{array}
\right.
\end{split}
,
\end{align}
where $t$ and $T$ are the numbers of current and total iteration respectively. Fig.~\ref{fig:Illu_DA} shows a profile of the training process adopting the technologies proposed herein.
We simply use the momentum based stochastic gradient decent as the optimizer and set the momentum as 0.9.
The learning rate changes following the cosine annealing from 0.01 to 0 during the training process.

\subsection{Cross-validation}

\begin{figure*}[t]
\centering
\includegraphics[width=1\linewidth]{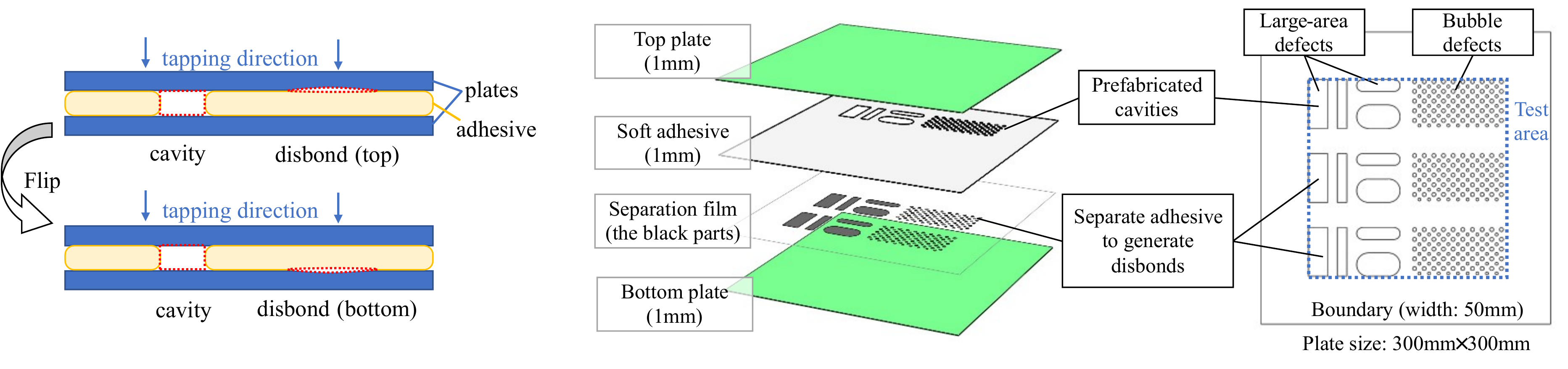}
\caption{Illustration of the specimens for acquiring data of the coin-tap signals. From the left, it is respectively the sketch map, exploded view, and the top view of the designed specimens for test.}
\label{fig:samples}
\end{figure*}

We adopt 3-fold cross-validation in all experiments.
Specifically, for the defect recognition in unitary material, the samples are randomly divided into three sections evenly.
The models are trained on two of the three sections, and tested on the remaining one.
In experiments, we report the mean accuracy and standard deviation of the three runs.
For the crossing material classification, we use all samples from the source domain and we conduct the cross-validation on the target domain.

\section{Data acquisition}

For validating the methods proposed herein, we constructed a benchmark dataset of the coin-tap test.
The followings illustrate some details of the data acquisition.

\subsection{Categorization of the near-surface defect}

Plates bonded with adhesive are the basic structure of widely-used composite laminates.
The most common types of defects on the bonded plates are cavity (lack of adhesive) and disbond.
According to this reality, we categorize the near-surface defect into 6 classes considering different defect types and shapes.
As shown in TABLE~\ref{TABLE:Categorization} and Fig.~\ref{fig:samples}, there is 7 categories of patterns of the sound signal in total including normal (N), large-area cavity (LC), bubble-like cavity (BC), large-area top disbond (LTD), bubble-like top disbond (BTD), large-area bottom disbond (LBD), bubble-like bottom disbond (BBD).
Concretely, the large-area defects are rectangles or rounded rectangles with a width of 10mm or 20mm.
The bubble-like defects are circular and with a diameter of 3mm or 4mm.
All the defects are at the depth range of 1mm to 2mm.
These defects are produced using the laser cutting machine before bounded together, so the shape and size are with good quality of manufacturing.

\subsection{Experimental specimens}
\begin{figure*}[t]
\centering
\includegraphics[width=0.9\linewidth]{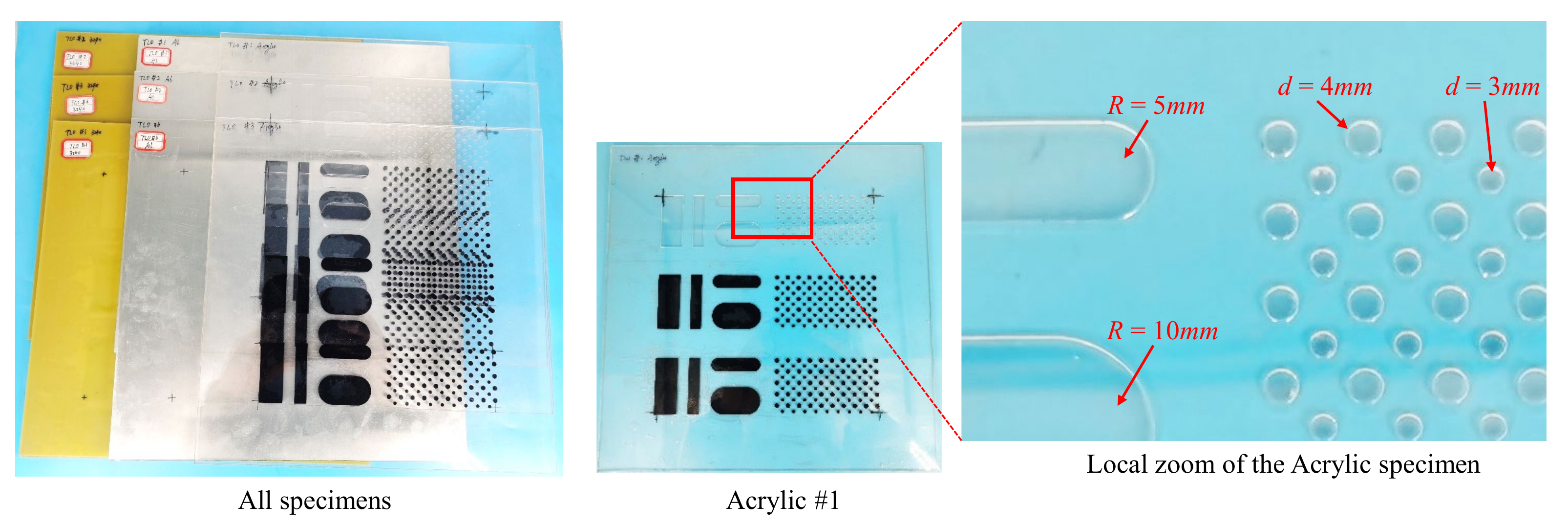}
\caption{Pictures of the real specimens.}
\label{fig:specimen}
\end{figure*}

The specimens are designed as square two-layer bonded plates, as shown in Fig.~\ref{fig:samples}.
The plates are bonded with 3M-VHB two-sided soft adhesive.
The thickness of both plates and the soft adhesive is $1mm$.
We use separation film to generate disbonds, and its thickness is thin enough to be negligible.
Totally, nine specimens of three different materials are prepared, including the Glass Fiber Reinforced Plate (GFRP), Acrylic, and Aluminum (Al). The selected three materials are representing composite materials, polymers, and metals, respectively, which could be a great range to validate the proposed method.
Pictures of the real specimens are in Fig.~\ref{fig:specimen}

The coin-tap tests are conducted in a square area of $200mm\times200mm$ at the center of the laminate.
As shown in the left of Fig.~\ref{fig:samples}, to make full use of the specimens, each specimen is tested from two sides.
This special design makes us be able to collect data easier.

\subsection{Test rig}
\begin{figure}[t]
\centering
\subfigure[schematic diagram]
    {
    \includegraphics[width=1\linewidth]{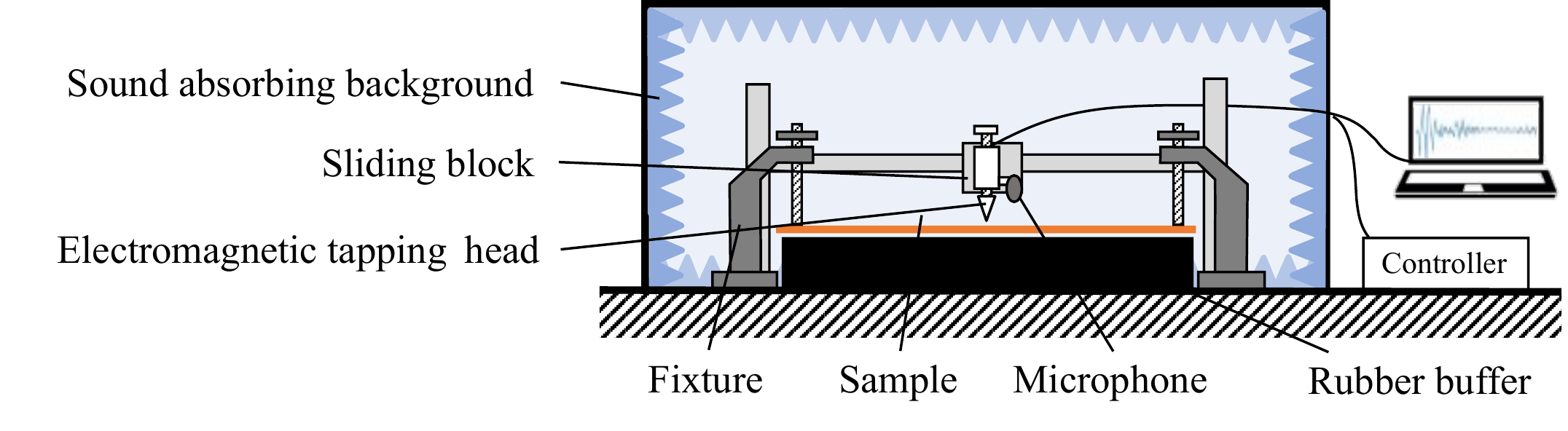}
    }
\subfigure[photographs]
    {
    \includegraphics[width=0.9\linewidth]{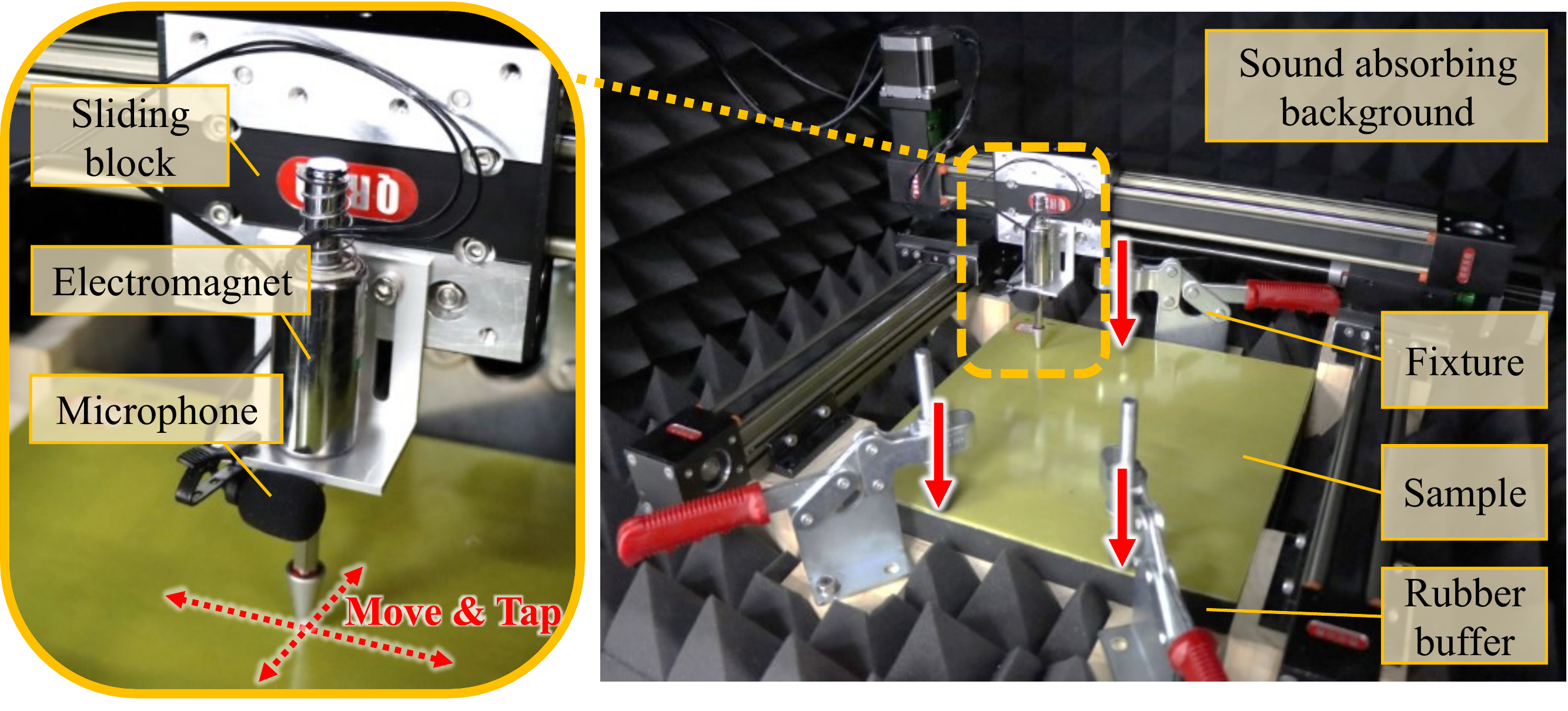}
    }
\caption{The test rig.}
\label{fig:test_rig}
\end{figure}

The test rig for acquiring data is comprised of a programmable automatic sliding block, an electromagnet for tapping, a microphone for receiving, a rubber buffer, and three fixtures for the specimens, as shown in Fig.~\ref{fig:test_rig}.
On each side of one specimen, the test rig is programmed to scan the test area of $200mm\times200mm$ with the stride of $2mm$, so we get 20402 sound signals.
Totally, the raw data includes 183618 signals comprised of 61206 signals respectively from each of the three materials.
The signals are recorded as .mp3 files at first, and it is then split and processed into data samples for analysis.
We adopt the sampling rate at 48kHz, which can meet the requirements of signal analysis~\cite{Lessmeier2016ConditionMO}.

\begin{figure}[t]
\centering
\includegraphics[width=1.02\linewidth]{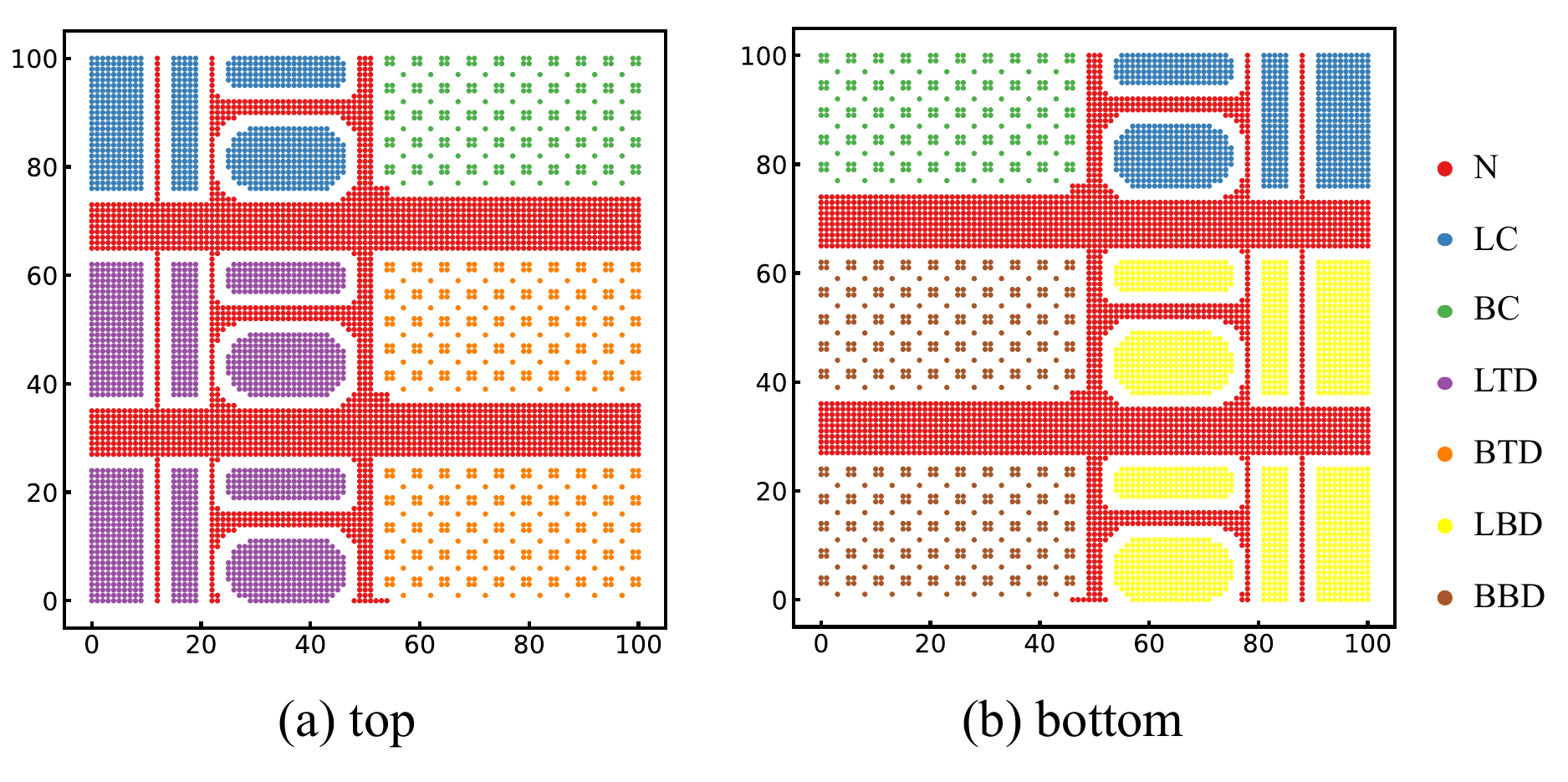}
\caption{Labels of the tapping map on the specimens. Color of the points denotes the class of the sound signal.}
\label{fig:labels}
\end{figure}

\subsection{Split and label the signals into a dataset}

We split the data into samples with the same length (2048), and then we labeled all the collected data by hands according to the tapping position and the prefabricated defects.
Fig.~\ref{fig:labels} presents a labeling map of all the points which is included in the dataset.
Note that some boundary point around the area of defects is removed to make the category of data clearer.

\begin{figure}[t]
\centering
\includegraphics[width=1\linewidth]{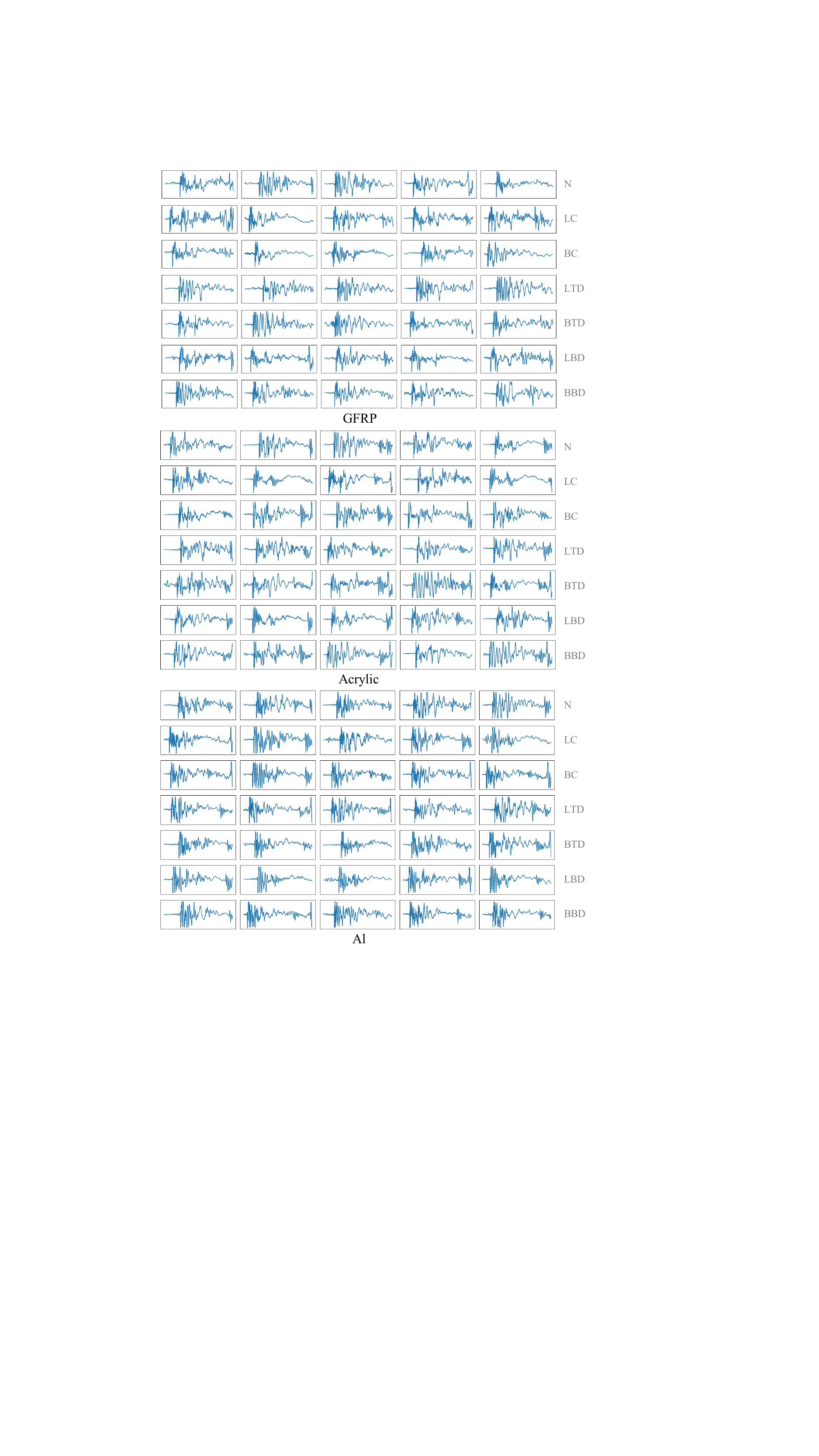}
\caption{Waveforms of the raw data collected in categories.}
\label{fig:raw_data}
\end{figure}

Theoretically, the sound and vibration induced via tapping are closely related to some mechanical properties such as the Young's modulus (elastic modulus) of materials.
For the materials (GFRP, Acrylic and Al) used herein, their elastic moduli differ by orders of magnitude.
Thus, we divided the data collected from different materials into different subsets, named by the material.
For each subset, the exact amount of data of different categories is shown in TABLE~\ref{TABLE:Categorization}.
In total, the dataset we constructed includes 101070 ($\sim$100k) signals of the coin-tap test.
Fig.~\ref{fig:raw_data} shows the waveforms of the raw data in categories.

\begin{table}[!t]
	\renewcommand{\arraystretch}{1.15}
    \caption{Categorization of the near-surface defect and sample numbers collected from one material.}
	\centering
	\label{TABLE:Categorization}
	\resizebox{\columnwidth}{!}{
        \setlength{\tabcolsep}{7pt}
		\begin{tabular}{c c c c}
			\hline\hline \\[-3mm]
			\multicolumn{1}{c}{Type} & \multicolumn{1}{c}{Shape \& Size} & \multicolumn{1}{c}{Short name} & \multicolumn{1}{c}{Number of samples}  \\ [0.6ex]
            \hline
			Normal & & N & 15978 \\
            \hline
            \multicolumn{1}{c}{\multirow{2}{*}{Cavity}}  & Large-area & LC & 4434\\
            \cline{2-4}
            & Bubble & BC & 1470\\
            \hline
			\multirow{2}{*}{Disbond (top)} & Large-area & LTD & 4434 \\
            \cline{2-4}
            & Bubble & BTD & 1470 \\
            \hline
            \multirow{2}{*}{Disbond (bottom)} & Large-area & LBD & 4434 \\
            \cline{2-4}
            & Bubble & BBD & 1470 \\
			\hline\hline
		\end{tabular}
	}
\end{table}

\section{Results and discussions}

\subsection{Defect classification with sufficient labeled data}

\begin{table*}[!t]
	\renewcommand{\arraystretch}{1.15}
	\caption{Results of defect classification using sufficient labeled data. (\%)}
	\label{TABLE:classification}
    \centering
	\resizebox{1.95\columnwidth}{!}{
        \setlength{\tabcolsep}{10pt}
		\begin{tabular}{c c c c c c c c}
			\hline\hline \\[-4mm]
			\multicolumn{2}{c}{\multirow{2}{*}{Method}} & \multirow{2}{*}{Input signal (the length)} & \multicolumn{3}{c}{Accuracies of} & \multirow{2}{*}{Mean} & \multirow{2}{*}{\makecell[c]{Inference \\ time (ms)}} \\
            \cline{4-6}
            & & & GFRP & Acrylic & Al \\ 
            \hline
			\multirow{4}{*}{Conventional} & RF~\cite{Breiman2001Random} & raw (2048) & 54.41$\pm$0.36 & 54.11$\pm$0.48 & 51.92$\pm$0.83 & 53.48 & \multirow{4}{*}{-} \\
            \cline{2-7}
            & \multirow{3}{*}{SVM~\cite{Cortes2004SupportVectorN}} & raw (2048) & 64.92$\pm$0.2 &65.67$\pm$0.2 & 63.4$\pm$0.18 & 64.66\\
            & & *MEF (29) & 63.14$\pm$0.51 & 62.07$\pm$0.34 & 59.81$\pm$0.46 & 61.67\\
			& & WPD + MEF (232) & 71.71$\pm$0.12 & 67.68$\pm$0.31 & 66.82$\pm$0.06 & 68.74\\
            \hline
            \multirow{4}{*}{Deep learning} & TICNN~\cite{Zhang2018ADC}& \multirow{4}{*}{raw (2048)} & 90.22$\pm$0.15 & 89.99$\pm$0.31 & 91.98$\pm$0.34 & 90.73 & 2.52\\
            & plain CNN (ours)& & 96$\pm$0.28 & 95.36$\pm$0.36 & 97.06$\pm$0.11 & 96.14 & 3.85\\
            & ResNet (ours)& & 94.64$\pm$0.28 & 93.43$\pm$0.14 & 96.22$\pm$0.24 & 94.76 & 5.24\\
            & DenseNet (ours)& & \textbf{96.16$\pm$0.15} & \textbf{95.49$\pm$0.29} & \textbf{97.3$\pm$0.21} & \textbf{96.32} & 16.71\\
			\hline\hline
            \multicolumn{5}{l}{* MEF is the abbreviation of manually extracted features.}
		\end{tabular}
	}
\end{table*}

\label{results_unitary}
Herein, the models are trained with sufficient well-labeled data which are from identical scenario with the test samples.
TABLE~\ref{TABLE:classification} presents the accuracies of different methods on three data subsets, respectively.
The first block in TABLE~\ref{TABLE:classification} lists the results from the classical machine learning methods, the random forest (RF)~\cite{Breiman2001Random} and support vector machine (SVM)~\cite{Cortes2004SupportVectorN}.
Especially for SVM, we adopted three different forms of input signals,
i) the raw signals of length 2048,
ii) 29 statistical features of time and frequency domain extracted by Ma \textit{et al.}~\cite{Ma2017LocallyLE},
and iii) the same manually extracted features but the signals are decomposed into 8 components by the wavelet packet (sym4) before feature extraction.
However, according to the result, it is implicit that SVM and RF can not recognize the raw signals well (the accuracies are 53.48\% and 64.66\%, respectively), and even with the help of WPD and manual feature extraction, it is not improved much (less than 5\%).
This reveals that the conventional manual-designed feature extraction is inadequate.

By contrast, deep CNNs learn to extract and organize the features automatically from the data.
The second block in TABLE~\ref{TABLE:classification} lists the results of deep CNNs.
We compared the three different architectures proposed herein and the TICNN model designed by Zhang \textit{et al.}~\cite{Zhang2018ADC}.
The number of trainable parameters of these four networks is designed to be approximate (TICNN: 66.2k, plain CNN \& ResNet: 58.03k, DenseNet: 57.46k), and their training processes are identical.
We adopted a mini-batch size of 32.
Each training process costs 50000 iterations.
It is clear from the results that CNNs significantly outperform the conventional methods by a considerable margin, the accuracy is improved by about 30\% and reached 96.32\%.
On the other hand, comparing CNNs for the tasks, the proposed networks perform better than TICNN.
This improvement could be attributed to more convolutional layers, removal of the fully-connected layers at the end (they are replaced with the global average pooling and $1\times1$ convolution), and meticulous selection of the hyper-parameters.

Therefore, with the support of these experimental results, a capable (the CNN model proposed herein presents really great accuracy) and easy (the model directly recognizes the defect from raw signals) intelligent coin-tap test is achieved.
In addition, among the three networks designed in this paper, DenseNet shows the best accuracy on all the data subsets of different materials.
Though DenseNet costs more time for inferencing the signals, 16.71ms is not a long time for processing a signal with the length of 2048 (it corresponds to 42.67ms).
Thus, we adopt the DenseNet architecture in the following studies about transfer learning.

\begin{table}[!t]
	\renewcommand{\arraystretch}{1.15}
	\caption{Classification accuracies (\%) after domain transfer under varying size of mini-batches.}
	\label{TABLE:batchsize}
    \centering
	\resizebox{1\columnwidth}{!}{
        \setlength{\tabcolsep}{5pt}
		\begin{tabular}{c c c c c c}
			\hline\hline \\[-4mm]
			\multirow{2}{*}{Task} & \multicolumn{5}{c}{Size of mini-batches} \\
            \cline{2-6}
            & 8 & 16 & 32 & 64 & 128 \\ 
            \hline
			*G$\rightarrow$Ac & 56.17$\pm$2.45 & 53.46$\pm$1.11 & 58.46$\pm$2.5 & 58.71$\pm$0.53 & 56.69$\pm$3.28 \\
            Ac$\rightarrow$G & 45.38$\pm$1.14 & 48.95$\pm$0.99 & 54.12$\pm$1.69 & 58.02$\pm$2.42 & 53.61$\pm$2.54 \\
            \hline
            Average & 50.78 & 51.21 & 56.29 & \textbf{58.37} & 55.15 \\
			\hline\hline
            \multicolumn{6}{l}{* G: GFRP, Ac: Acrylic.}
		\end{tabular}
	}
\end{table}

\begin{table*}[!t]
    \renewcommand{\arraystretch}{1.15}
	\caption{Comparisons of classification accuracies of different methods in cases of transfer application. (\%)}
	\label{TABLE:domain_adaptation}
    \centering
	\resizebox{1.9\columnwidth}{!}{
        \setlength{\tabcolsep}{12pt}
		\begin{tabular}{c c c c c c c}
			\hline\hline \\[-4mm]
			\multirow{2}{*}{Situation} & \multirow{2}{*}{Methods} & \multicolumn{3}{c}{Transfer applications acrossing different materials} & \multirow{2}{*}{\makecell[c]{Mean \\ accuracy}}\\
            \cline{3-5}
            & & *G,Al $\rightarrow$ Ac & Ac,Al $\rightarrow$ G & G,Ac $\rightarrow$ Al &  \\ 
            \hline
			\multirow{6}{*}{\makecell[c]{No data from the target \\ task is labeled}} & *SVM without TL~\cite{Cortes2004SupportVectorN} & 24.96 & 34.64 & 21.34 & 26.98 \\
            & TCA~\cite{Pan2011DomainAV} & 33.62 & 34.24 & 33.28 & \underline{33.71}\\
            & JDA~\cite{Long2013TransferFL} & 33.20 & 35.00 & 32.74 & 33.65\\
            & CORAL~\cite{Sun2016ReturnOF} & 33.01$\pm$0.86 & 34.53$\pm$0.61 & 31.95$\pm$0.23 & 33.16\\
            \cline{2-6}
            & DenseNet without TL & 49.18 & 53.44 & 40.74 & 47.79 \\
            & DenseNet + DA + PLL & 60.51$\pm$2.61 & 62.22$\pm$1.51 & 56.49$\pm$2.10 & \textbf{\underline{59.74}}\\
            \hline
            \multirow{3}{*}{\makecell[c]{2\% data labeled}} & *SVM without TL~\cite{Cortes2004SupportVectorN} & 50.62 & 51.01 & 48.56 & 50.06 \\
            & DenseNet without TL & 65.87 & 68.20 & 67.95 & 67.34 \\
            & DenseNet + DA + PLL & 74.09$\pm$0.22 & 72.94$\pm$1.53 & 72.51$\pm$0.15 & \textbf{\underline{73.18}} \\
            \hline
            \multirow{3}{*}{\makecell[c]{5\% data labeled}} & *SVM without TL~\cite{Cortes2004SupportVectorN} & 46.78 & 49.18 & 46.69 & 47.55\\
            & DenseNet without TL & 77.00 & 77.20 & 76.46 & 76.89 \\
            & DenseNet + DA + PLL & 81.42$\pm$0.55 & 83.13$\pm$0.72 & 81.70$\pm$0.58 & \textbf{\underline{82.08}}\\
            \hline
            \multirow{3}{*}{\makecell[c]{10\% data labeled}} & *SVM without TL~\cite{Cortes2004SupportVectorN} & 60.69 & 63.71 & 60.15 & 61.52\\
            & DenseNet without TL & 82.73 & 84.84 & 85.15 & 84.24 \\
            & DenseNet + DA + PLL & \underline{86.71}$\pm$0.09 & \underline{89.36}$\pm$0.38 & \underline{88.93}$\pm$0.26 & \textbf{\underline{88.33}}\\
			\hline\hline
            \multicolumn{6}{l}{* TL is the abbreviation of transfer learning, DA: domain adaptation, PLL: pseudo-label learning, G: GFRP, Ac: Acrylic.} \\
		\end{tabular}
	}
\end{table*}

\subsection{Transfer to conditions without sufficient labeled data}

To apply the models into scenarios that can hardly get enough well-labeled data for training, we take the crossing material application of the CNN models as an example, and conduct studies on the proposed transfer learning strategies as follows.
Throughout the experiments in this part, we assume that data from one of three materials as the target task without label.
The intelligent models are trained mainly on data of other one or two materials and then applied to the target task.

\subsubsection{Selection of the key parameters}

To achieve better performance of the adopted domain adaptation strategy, we study two key parameters in calculating MMD: i) the bandwidth $\sigma$ of the Gaussian kernel; ii) the size of the mini-batches during network training.

\begin{figure}[t]
\centering
\includegraphics[width=0.95\linewidth]{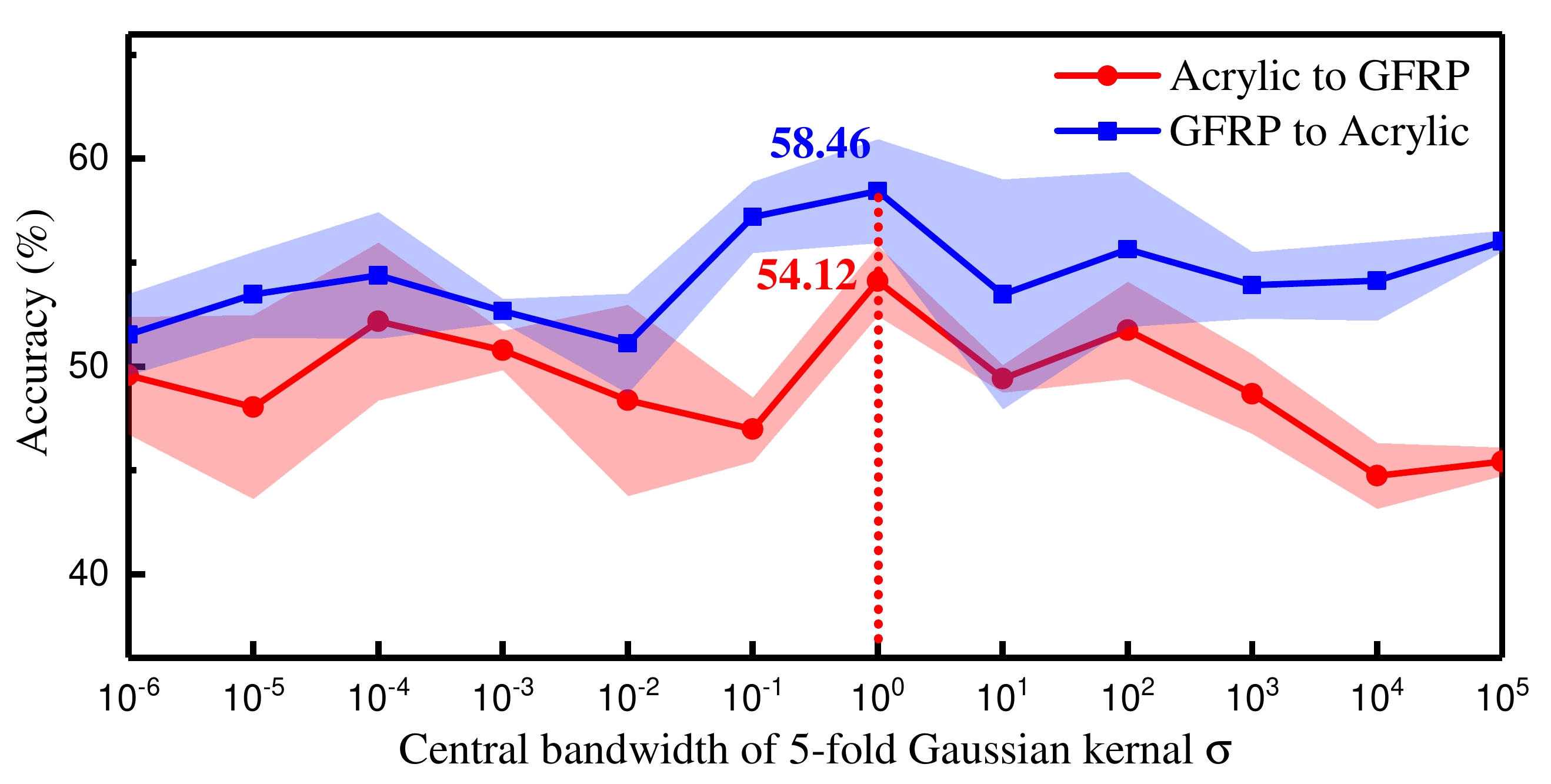}
\caption{Classification accuracies of sound signal recognition under varying bandwidth $\sigma$.}
\label{fig:bandwidth}
\end{figure}

The central bandwidth $\sigma$ of the 5-fold Gaussian kernel is a crucial parameter of the kernel function $k(\cdot,\cdot)$, which is used to map inputs into the RKHS implicitly.
According to Eq.~(\ref{fun:MMD}), (\ref{fun:kernel}) and the analysis by Yang \textit{et al.} in figure 1 of their paper~\cite{Yang2020APK}, $\sigma$ has significant impact on the calculation of MMD.
Hence, we carry out experiments adopting different $\sigma$.
Results are shown in Fig.~\ref{fig:bandwidth}.
Since the two transfer experiments are mutual, their general trend with $\sigma$ are similar, and we get the best accuracy at $\sigma=1$.

As Eq.~(\ref{fun:MMDt}), MMD is defined as the discrepancy of two distributions.
Since we can hardly use this analytical expression in practice, it is rewritten as a sampling estimation of Eq.~(\ref{fun:MMD}).
This estimation is made according to one mini-batch of samples, thus, the estimation becomes indistinct if the mini-batch size is too small.
Therefore, it is suggested to use greater mini-batch size to estimate the MMD better.
However, according to the common experience in the field of deep neural network training, large mini-batch sizes may limit the network's potential to gain better performance because the randomness of the training process is suppressed.
To make a choice of the size of the mini-batches, we train and test the models with different values of it.
The results are in TABLE~\ref{TABLE:batchsize}.
As expected, the models trained with small mini-batches exhibit poor performance owing to the inaccurate estimation of MMD.
The models trained with large mini-batches perform better, but the models trained with oversized mini-batches do not yield the best result.
Thus, the optimal mini-batch size of 64 is adopted in the following experiments.

\subsubsection{Comparison of different method in transfer application}

We consider four situations of the transfer application to scenarios without sufficient labeled data: i) No data from the target task is labeled; ii) 2\% data labeled; iii) 5\% data labeled; iv) 10\% data labeled. Specifically, we regard the coin-tap test on different materials as the target scenarios for application.
Especially in the first situation, different methods are compared.
Three conventional transfer learning methods are considered herein, including transfer component analysis (TCA)~\cite{Pan2011DomainAV}, Joint Distribution Adaptation (JDA)~\cite{Long2013TransferFL} and CORAL~\cite{Sun2016ReturnOF} with K-nearest neighbors as the classifier.
They use manually extracted features, that 232 statistical features are extracted for one signal, the same as section~\ref{results_unitary}.
The results are listed in TABLE~\ref{TABLE:domain_adaptation}.
We report the mean accuracy and the standard deviation obtained from the cross-validations\footnote{Note that cross-validation is not applicable to some of the methods, such as TCA, JDA and methods without transfer learning. In fact, their accuracies exhibit little randomness, so we just use all source/target data and report the mean accuracy of three runs.}.
It is clear from the results that, firstly, the DenseNet (both with or without transfer learning technologies) outperforms conventional machine learning methods (26.98\% vs. 59.74\% or 61.52\% vs. 88.33\%) because of better feature extraction from the raw signals.
Further, by adopting the domain adaptation and pseudo label learning strategy proposed herein, the CNN models can learn more transferable knowledge of the defect signals from unlabeled data of the target domain.
These transfer learning strategies improve the accuracy in all four situations, especially, for the case with no labeled data from the target task, they got the biggest improvement as 11.95\% (from 47.79\% to 59.74\%).
Besides, for the case with 10\% data labeled, the model achieves the best accuracy at 88.33\%, which could meet the requirement of some real applications.

\begin{figure*}[t]
\centering
\includegraphics[width=0.8\linewidth]{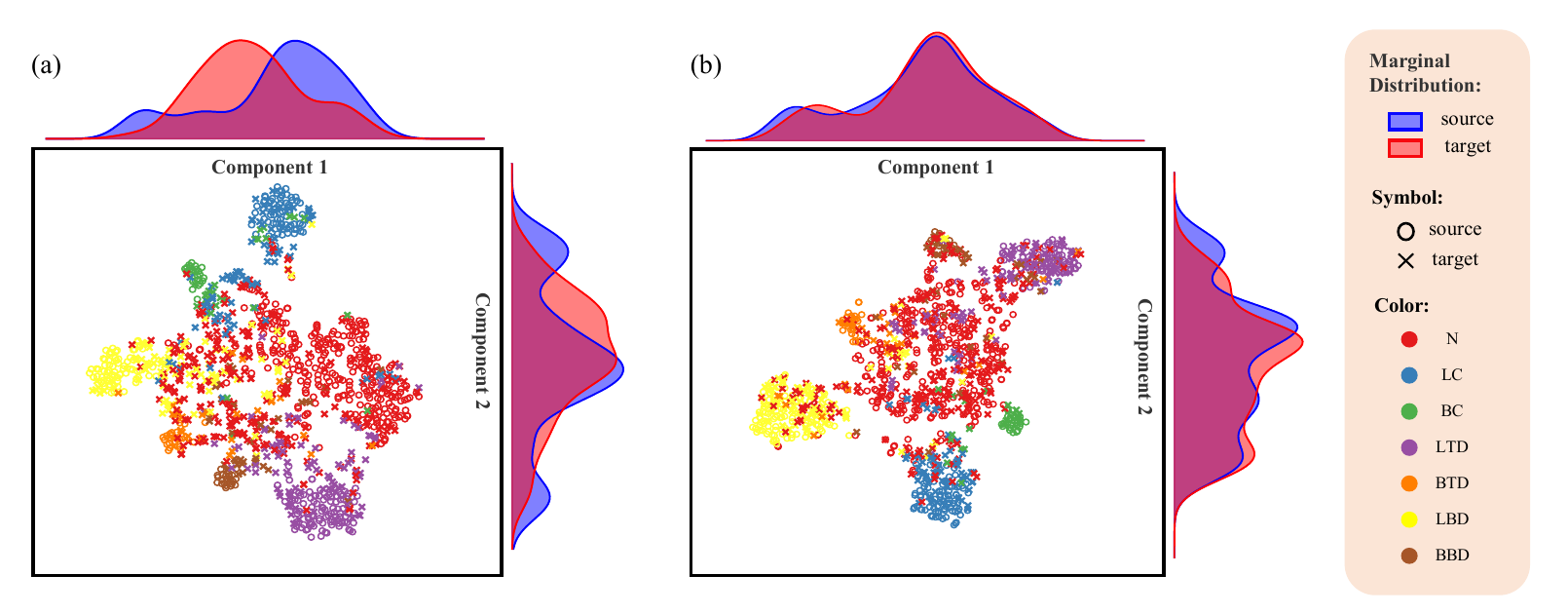}
\caption{The 2-D t-SNE embedding on the task of Ac,Al$\rightarrow$G by: (a) DenseNet without transfer learning (accuracy: 53.44\%), (b) DenseNet + domain adaptation + pseudo label learning (64.11\%). For each component, its distribution is estimated by kernel density estimation and attached outside the coordinate axis.}
\label{fig:embedding}
\end{figure*}

\begin{figure*}[t]
\centering
\includegraphics[width=1\linewidth]{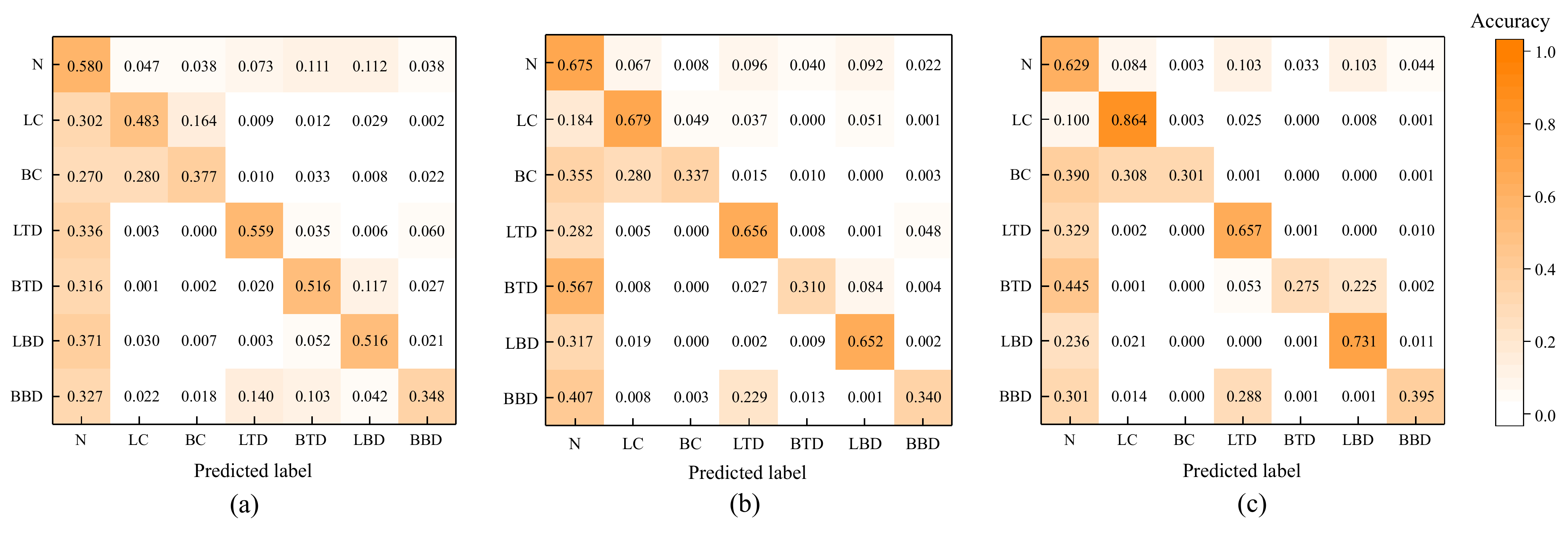}
\caption{The confusion matrix of the defect recognition on the task of Ac,Al$\rightarrow$G by: (a) DenseNet without transfer learning (accuracy of this instance: 53.44\%), (b) DenseNet + domain adaptation (62.34\%), (c) DenseNet + domain adaptation + pseudo label learning (64.11\%).}
\label{fig:confusion}
\end{figure*}

We present more discussions about the results here to understand the characteristics of the adopted technologies better.
Considering the task of Ac,Al$\rightarrow$G in the situation of no labeled data for instance, the 2-D t-SNE embedding of the different methods is shown in Fig.~\ref{fig:embedding}.
Focusing on the distribution of samples of the source/target domain, the network trained with domain adaptation by minimizing MMD can narrow the gap between the probability distributions of the source and target domains.
Thus, the accuracy is improved.
However, even if the minimization of the MMD is achieved, a considerable number of samples are still incorrectly classified, which could be a limitation.
Besides, the confusion matrixes shown in Fig.~\ref{fig:confusion} reveal the pros and cons of pseudo label learning.
Comparing Fig.~\ref{fig:confusion}(c) with \ref{fig:confusion}(a) or \ref{fig:confusion}(b), it shows that the pseudo label learning makes networks more 'stubborn' whether the recognition is correct or not.
This is because the strategy of pseudo label learning amplifies the knowledge learned.
It improves the accuracies of categories that are already well-learned (N, LC, LTD, and LBD), but is not as useful to the others (bubble-like defects), which could be a limitation.

\section{Conclusion}

This work developed a deep learning based intelligent coin-tap test for defect recognition.
The proposed method is exemplified via laboratory experiments using specimens with prefabricated millimeter-level defects.
For one thing, in scenarios with sufficient well-labeled data from the identical task, the CNN models constructed herein outperform conventional methods with a considerable margin (the accuracy is improved by about 30\% and reaches 96.32\%).
For another thing, considering situations that can hardly get plenty of labeled data, we carried out studies about model transfer across different similar tasks, specifically, we tried different materials as an example here.
The result shows that, the intelligent models are still effective in the transfer application, especially when adopting domain adaptation and pseudo label learning strategies proposed herein (the accuracy is improved by over 25\% and reaches 88.33\% using 10\% labeled data).
This means, with these transfer learning strategies, it becomes possible to apply the model into scenarios with none or little (less than 10\%) labeled data, which may greatly promote this advanced method to real application.

\bibliographystyle{IEEEtranTIE}
\bibliography{reference}\ 

\end{document}